\documentclass[
    ,final            
  ]
  {aipproc}

\layoutstyle{8x11single}

\begin{document}

\title{Cataclysmic Variables: Eight Breakthroughs in Eight Years}


\author{Christian Knigge}{
  address={School of Physics \& Astronomy, University of Southampton, Southampton SO17 1BJ, UK}
}



\begin{abstract}
The last few years have seen tremendous progress in our understanding 
of cataclysmic variable stars. As a result, we are finally developing
a much clearer picture of their evolution as binary systems, the
physics of the accretion processes powering them, and their relation
to other compact accreting objects. In this review, I will highlight
some of the most exciting recent breakthroughs. Several of these have
opened up completely new avenues of research that will probably lead
to additional major advances over the next decade.
\end{abstract}

\maketitle


\section{Introduction}

The study of cataclysmic variables (CVs) -- close binary systems
containing an accreting white dwarf (WD) primary -- has been undergoing
a renaissance over the last few years. As also recently
noted by Paul Groot \cite{groot2010}, the field had experienced a 
boom in the 80s and early 90s, but then seemed to suffer a bit of a
slump. This seems to have been caused partly by the need to shift
focus from what 
used to be a mostly ``object-centered'' view of the field to one that
is more ``population-centered''. As I will try to show in this review,
this slump is most definitely behind us. In fact, the last few years
have seen a series of breakthroughs that are dramatically improving
our understanding of CV evolution, accretion physics and the
connection between CVs and related systems, such as accreting neutron
stars (NSs) and black holes (BHs). Let me start, however, by providing
some context for these advances.

\section{Cataclysmic Variables: A Primer}


\subsection{The Physical Structure of CVs}

CVs are semi-detached close binary systems in which a a WD accretes
material from a Roche-lobe-filling secondary. In most known CVs, the
secondary is (almost) a main-sequence (MS) star, and the transfer of
mass from the secondary to the WD happens via an accretion
disk. The orbital periods of CVs are typically between 75~min 
and 6~hrs, although there are exceptional systems -- usually with
evolved or compact donor stars -- with periods outside this range. 

\subsection{CV Evolution: The Standard Model}

\begin{figure}[t]
  \includegraphics[height=.35\textheight]{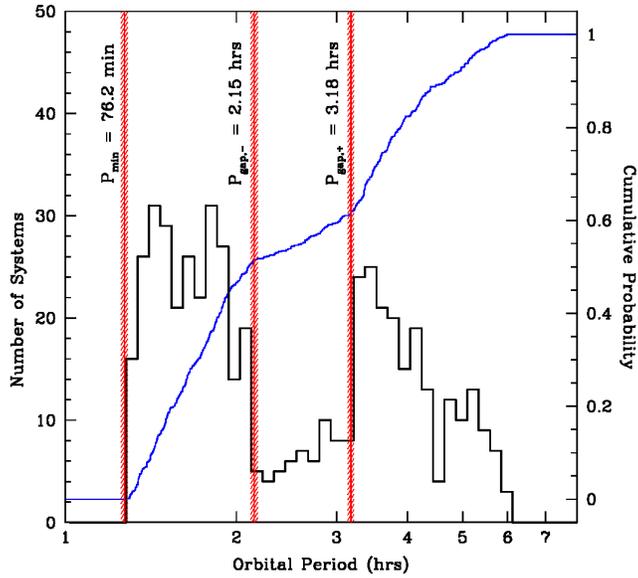}
  \caption{ Differential and cumulative orbital period distribution of
  CVs, based on data taken from Edition 7.6 of the Ritter \& Kolb
  catalogue \cite{ritter2003}. Estimated values for the minimum period and the
  period gap edges are shown as vertical lines. The shaded regions
  around them indicate our estimate of the errors on these
  values. Figure reproduced by permission from
  Ref. \cite{knigge2006}.}
\label{fig:pdist}
\end{figure}

Most of our attempts to understand the secular evolution of CVs have
been driven by a single plot. This plot is the orbital period
distribution of CVs, a fairly recent version of which is shown in
Figure~\ref{fig:pdist}. This distribution exhibits two key
features. First, there is an 
obvious deficit of CVs in the period 
range between 2~hrs and 3~hrs; this is the famous CV ``period
gap''. Second, there is a sharp cut-off near $P_{min} \simeq 80$~min;
this is the so-called ``period minimum''.

The standard model of CV evolution that attempts to explain these
features was conceived almost 30~years
ago \cite{robinson1981,rappaport1982,rappaport1983,spruit1983}. Briefly, 
stable mass 
transfer in a CV containing an initially unevolved MS donor star is
only possible in the presence of angular momentum loss (AML) from the
system. This (initially) shrinks the binary orbit and keeps 
the Roche lobe in contact with the mass-losing and also shrinking
secondary star. Thus CVs (initially) evolve from long to short orbital
periods.  

According to the standard model, above the period gap,
the AML mechanism that drives CV evolution is {\em magnetic braking}
(MB), i.e. a magnetized stellar wind from the donor star. MB can be
quite strong and thus drives a fairly high mass-transfer rate. In
fact, MB-driven mass loss takes place at a rate that is comparable to
the secondary's thermal time-scale. As a result, the donor is driven slightly out of 
thermal equilibrium: its radius cannot adjust quite fast enough to its
ever-decreasing mass. Thus the donor always slightly bloated relative
to an equal-mass, isolated MS star.

The combination of semi-detached geometry and Kepler's third law
implies that there is an (almost) unique, monotonic relationship
between the orbital period of a CV and the mean density (and hence
mass) of its secondary. As it turns out, the upper edge of the period
gap corresponds to roughly the point where the donor is expected to
lose its radiative core and become fully convective. 
The standard model therefore posits that this transition will be
accompanied by the cessation of MB. The justification for this is
that the magnetic fields of low-mass stars are often assumed to be
anchored in the tachocline, i.e. the transition region between the
radiative core and the convective envelope. 
With MB gone, the only remaining AML mechanism
is gravitational radiation (GR). This operates at a much slower
rate and is unable to sustain the same high mass-loss rate from the donor
star. The donor therefore shrinks closer to its thermal equilibrium
radius, but in doing so loses contact with the Roche lobe
completely. The upper edge of the period gap thus marks the beginning
of a detached phase for CVs. 

Evolution through the gap is still from long to short periods, 
as GR continues to slowly shrink the orbit. Contact is
eventually reestablished when the size of the Roche lobe is equal to
that of a MS star in thermal equilibrium. This marks the lower
edge of the period gap in the standard model. Mass transfer then
resumes, and the system once again evolves as an active CV to even
shorter periods. But this phase of evolution cannot continue
indefinitely either. In particular, brown dwarfs have an inverted
mass-radius relationship, so donors with masses 
well the hydrogen-burning limit may be expected to {\em grow} 
in size in response to mass loss. Since the period-density 
relationship still applies, the orbital period of a CV with
a sub-stellar companion must then also increase. Thus at some 
point during the transition of the donor from a very low-mass MS star 
to a strongly sub-stellar object, the system must reach a minimum
period.This, then, is the standard explanation for the sharp
cut-off in the CV period distribution.
\footnote{It is actually easy to show that $P_{min}$ occurs exactly 
when the effective mass-radius index of the donor along the
evolution track reaches $\zeta = 1/3$ (see, for example,
Ref. \cite{knigge2006}). It should also be noted that this 
type of ``period bounce'' does not {\em necessarily} have to be
associated with a stellar to sub-stellar transition. Any low-mass star
with a deep convective envelope will grow in radius (with
$\zeta \simeq -1/3$) if exposed to mass loss on a time scale much
shorter than its own thermal time scale.} 

This ``disrupted magnetic braking'' picture has
dominated thinking about CV evolution ever since 
its inception. However, it is fair to say that, until recently, is had
remained largely untested. Its ability to explain the 
period gap and the period minimum is certainly impressive, but then
explaining these features is what the model was designed to
do. In fact, it has been known for some time that some other, 
quantitative predictions of the model appear to be in conflict with
observations. For example, the model predicts a shorter-than-observed 
minimum period, as well as too few long-period CVs compared to
short-period ones, even when allowing for selection 
effects \cite{pretorius2007}. Does this mean the standard model is
fundamentally wrong? Or does it just need ``tweaking'', such as
allowing for some extra AML in addition to GR acting below the period
gap (e.g. \cite{patterson1998})? Or is the standard model actually correct -- 
are the apparent conflicts with observations just due to our
inability to properly model CVs and the selection biases that affect
them? 

\section{Eight Breakthroughs in Eight Years}

In the following sections, I will describe what I consider to be 
eight of the most important advances in CV research over the last
decade. Several of them represent the first proper tests of the basic
evolutionary scenario outlined above, but there have also been
key breakthroughs in our understanding of accretion physics and of the
connection between CVs and other classes of compact accreting objects. 

\subsection{Breakthrough I: Disrupted Angular Momentum Loss at the
Period Gap}

As noted above, one of the key goals in the design of the standard
model was to provide a cogent explanation for the existence of the
period gap. However, it is remarkably difficult to properly test the
idea that the gap is caused specifically by a disruption of AML -- as
opposed to, for example, the presence of distinct populations above and
below the gap (e.g. \cite{andronov2003}). However, there is one key prediction of the
model that can, in principle, be tested: if the standard model is correct,
donors just above and below the gap should have identical masses, but
different radii. After all, the donors above the gap have been
significantly inflated by mass loss, while CVs below the gap
have just emerged from a detached phase with their donors in thermal
equilibrium.

\begin{figure}[t]
  \includegraphics[height=.3\textheight]{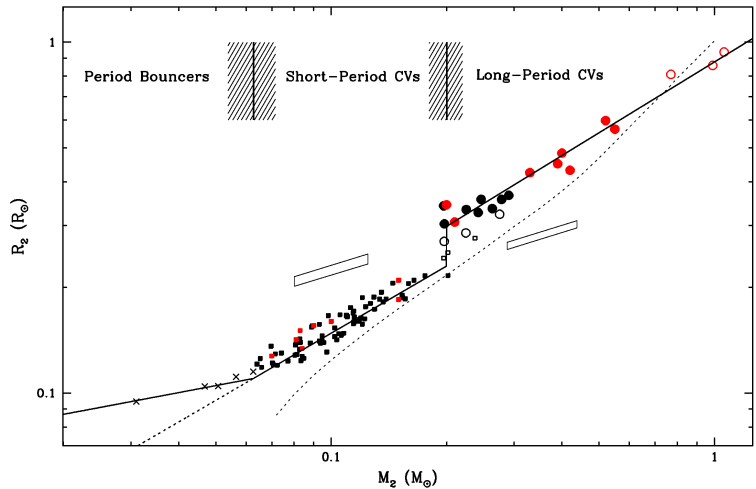}
  \caption{The mass-radius relation of CV donor stars, based on the
  data presented in \cite{patterson2005}. Superhumpers are shown
  in black, eclipsers in red. Filled 
  squares (circles) correspond to short-period (long-period) CVs,
  crosses to likely period bouncers. The 
  parallelograms illustrate typical errors. 
  Open symbols correspond to systems in the period gap or likely
  evolved systems. The solid lines show the optimal broken power-law 
  fit to the data. The
  dotted line is a theoretical mass-radius relation for MS stars
  \cite{baraffe1998}. Figure
  reproduced and adapted by permission from Ref. \cite{knigge2006}.} 
\label{fig:mr}
\end{figure}

In 2005, Joe Patterson showed for the first time that this fundamental
prediction is correct \cite{patterson2005}. Over almost two decades of
painstaking work, he and his ``Center for Backyard Astronomy''
collaborators collected a vast amount of observational data on ``superhumps''
in CVs and showed that these observations can be calibrated to yield mass 
ratios for these systems. These mass ratios, in turn, can be used to
obtain estimates of the corresponding donor masses and radii. 
He then combined these with similar estimates obtained for eclipsing
CVs (such estimates are more precise, but available for 
far fewer systems) and put together the mass-radius relationship for
CV donor 
stars shown in Figure~\ref{fig:mr}.\footnote{Actually, the figure here
is from Ref. \cite{knigge2006}, but the data are based entirely on
Patterson's compilation in Ref. \cite{patterson2005}.}

The main result is immediately apparent: there is a clear
discontinuity in donor radii at $M_2 \simeq 0.2 
M_{\odot}$ that also cleanly separates long-period from short-period
systems. In fact, donors in systems just below the period gap have
radii consistent with ordinary MS stars of equal mass, while donors
just above the gap have radii that are inflated by $\simeq 30\%$. All
of these findings are exactly in line with the basic predictions of
the disrupted MB model.

Before moving on, it is worth emphasizing that Figure~\ref{fig:mr} alone
cannot tell us the exact nature of the disruption in AML responsible
for the period gap. In particular, {\em any} significant reduction of
AML at $P \simeq 3$~hrs will produce a period gap and a 
discontinuity in the donor mass-radius relationship. Without further
modelling, the data cannot tell us if the AML above the gap has the 
strength expected for MB, nor if MB ceased completely or was merely 
somewhat suppressed at the upper gap edge. However, Figure~\ref{fig:mr}
is extremely strong evidence for the basic idea of a disruption in 
AML at the upper gap edge. It thus represents a tremendously important
advance.

\subsection{Breakthrough II: The Existence of CVs with Brown Dwarf Secondaries}

A second key prediction of the standard model of CV evolution is that
most CVs should already have evolved past the period minimum, i.e. they
should be ``post-period-minimum systems'' or ``period
bouncers''. In fact, the standard model predicts that about 70\% of
present day CVs should be period bouncers, with all of these
possessing sub-stellar donor stars (e.g. \cite{kolb1993}). It was
therefore quite disconcerting that, until recently, only a handful of
{\em candidate} period bouncers were known. In particular, there was
not even one CV with a well-determined donor mass below the
Hydrogen-burning limit.

This situation has finally changed, thanks to the population-centered
approach mentioned in the introduction. Over the course of several
years, Paula Szkody and collaborators have produced a new sample of
$\simeq$200 CVs from the Sloan Digital Sky Survey 
(SDSS; \cite{szkody1,szkody2,szkody3,szkody4,szkody5,szkody6}).
This sample has a much deeper effective magnitude limit
than previous ones and is therefore much more sensitive to the very
faint CVs near and beyond $P_{min}$. Crucially, the new SDSS sample
included several new {\em eclipsing} candidate period bouncers, for
which component masses could be determined geometrically by careful
modelling of high-quality eclipse observations. 

\begin{figure}[t]
  \includegraphics[height=.19\textheight]{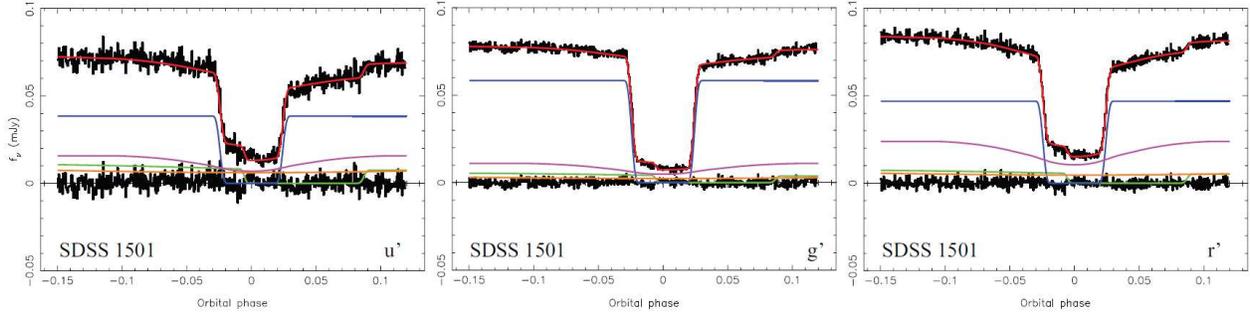}
  \caption{Model fits to the phase-folded u$^\prime$, g$^\prime$ and
  r$^\prime$ light curves of 
  SDSS J1501. The data (black) are shown with the fit (red) overlaid
  and the residuals plotted below (black). Below are the separate
  light curves of the WD (blue), bright spot (green), accretion disc
  (purple) and the secondary star (orange). Figure adapted and
  reproduced by permission from Ref. \cite{littlefair2008}.}
\label{fig:eclipse}
\end{figure}

Such eclipse analyses have been carried out by Stuart Littlefair and
collaborators \cite{littlefair2006,littlefair2008} and have so far
yielded three significantly sub-stellar donor mass estimates. An
example of a light curve and model fit for one of these systems --
SDSS J1501, whose donor has a mass of $M_2 = 0.053 \pm 0.003
M_{\odot}$ -- is shown in Figure~\ref{fig:eclipse}.

The definitive detection of CVs with sub-stellar donors is a huge
result. It does not prove that the 
standard model is correct -- it is still far from clear, for example,
whether there are enough of these systems in the Galaxy to be
consistent with theoretical predictions. However, it does confirm the
fundamental idea that (at least some) systems survive the
stellar-to-substellar transition of their secondaries, while remaining
active, mass-transferring CVs.

\begin{figure}[t]
  \includegraphics[height=.35\textheight]{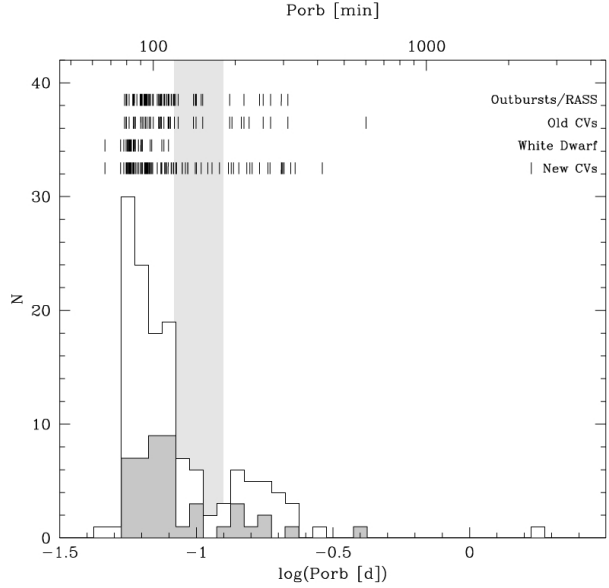}
  \caption{The period
  distribution of SDSS CVs, divided into 45 previously known
  systems (old SDSS CVs, grey) and 92 newly identified CVs (new SDSS
  CVs, white). Superimposed are tick marks indicating the individual
  orbital periods of the old and new SDSS CVs, those of SDSS CVs
  showing outbursts, those of SDSS CVs detected in the ROSAT 
  All-Sky Survey, and those of SDSS CVs which reveal
  the WD in their optical spectra. Figure  adapted and reproduced by
  permission from Ref. \cite{boris2009}).}
\label{fig:spike}
\end{figure}

\section{Breakthrough III: The Discovery of the Period Spike}

Another long-standing prediction of the basic evolution scenario
for CVs is that there should be a ``period spike'' at the 
minimum period (e.g. \cite{kolb1999}). More specifically, the orbital period
distribution of any sufficiently deep sample should show at least a
local maximum near $P_{min}$. This prediction is easy to understand:
the number of CVs we should expect to find in any period interval is
proportional to the time it takes a CV to cross this
interval, $N(P) \propto \dot{P}^{-1}$. But $\dot{P}(P_{min}) = 0$, 
so the period interval including $P_{min}$ should contain an unusually 
large number of systems. This is a critical prediction, since it
follows directly from the idea that $P_{min}$ marks a change in the
direction of evolution for CVs.

Until recently, no CV sample or catalogue showed any sign of the
expected period spike (e.g. Figure~\ref{fig:pdist}). 
However, CVs near $P_{min}$ are very faint, so it was recognized
that this could just be due to a lack of depth in these
samples \cite{barker2003}.

Here again, the new population-focused emphasis mentioned above,
implemented via several years of hard work, has yielded a definitive
answer. In fact, the breakthrough in this area was again driven by
the SDSS CV sample, and, specifically, by a long-term effort led by Boris
G\"{a}nsicke to obtain precise orbital periods for these
systems \cite{boris2009}. Figure~\ref{fig:spike} shows 
the resulting period distribution for the SDSS CVs. The period spike 
near $P_{min}$ is clearly visible.

The existence of the period spike does not necessarily imply that the
standard model is quantitatively correct. In particular, it does not
mean that 
AML below the gap is driven solely by GR. In fact, the location of the
spike at $P_{min} \simeq 82$~min is even further from the 
prediction of the standard model ($P_{min} \simeq 65
-70$~min; e.g. \cite{kolb1993}) than
previous empirical estimates (which put $P_{min} \simeq
75$~min; e.g. \cite{knigge2006}). Stronger-than-GR AML below the gap
may be required to 
reconcile this discrepancy between theory and observations. However, 
the discovery of the period spike in the SDSS sample provides
convincing evidence for the fundamental prediction that CVs actually
undergo a period {\em bounce} at $P_{min}$. As such, it represents a
massive step forward in our understanding of CV evolution. 

\section{Breakthroughs IV and V: Reconstructing CV Evolution from
Primaries and Secondaries}

\begin{figure}[t]
  \includegraphics[height=.245\textheight]{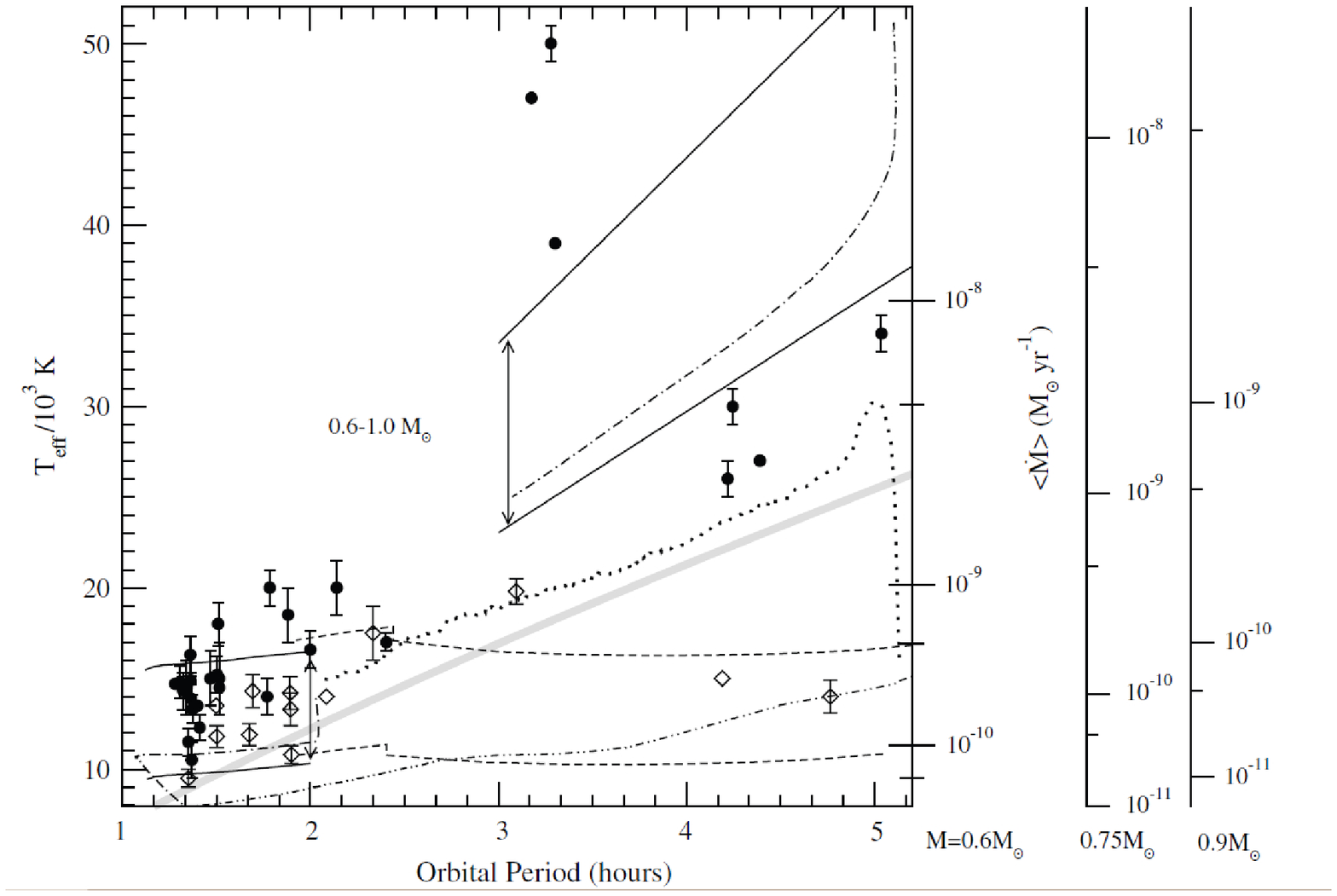}
  \includegraphics[height=.245\textheight]{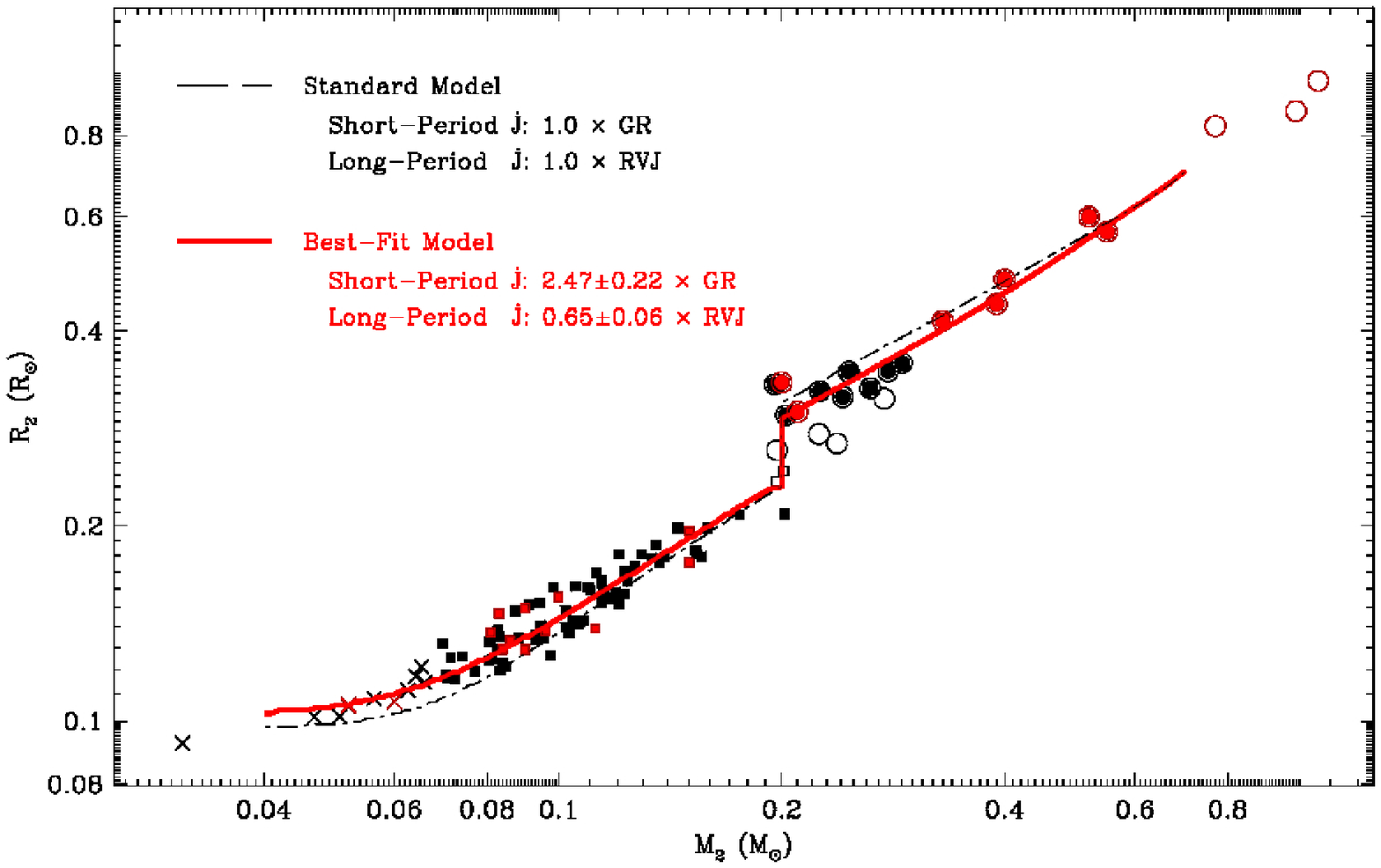}
  \caption{{\em Left Panel:} Reliable $T_{eff}$ measurements for
  CVs. An approximate mapping to
  $\dot{M}$ is shown on the right vertical scale
  assuming $M_{WD} = 0.75 M_{\odot}$, $0.6 M_{\odot}$, or $0.9
  M_{\odot}$. Several sets of predicted temperatures are also
  indicated: an empirical relation (\cite{patterson1984}, thick grey
  line), traditional MB 
  \cite{howell2001}, dot-dashed line and between solid lines), 
  reduced MB (\cite{andronov2003}, dot-dot-dash line; 
  \cite{ivanova2003}, dotted line), pure GR (between dashed lines). 
  Figure reproduced by permission from
  Ref. \cite{townsley2009}. {\em Right Panel:} Model fits to the
  observed CV donor mass-radius. The thin 
  dashed line is the relationship predicted by the standard
  model, the thick solid line shows the optimal fit achieved by 
  varying the strength of AML above and below the gap. 
  Figure from Ref. \cite{knigge2010}.}
\label{fig:reconstruct}
\end{figure}

The advances described above have finally provided us with strong
evidence that our basic ideas about CV evolution are at least
qualitatively correct. But does the standard model agree {\em
quantitatively} with observations? What is the strength of MB above
the period gap? Is GR really the only AML mechanism acting below the
gap? These issues are central not only to CVs, but to virtually all
types of close binaries, since  AML via MB and/or GR are thought to
drive the evolution of these systems also.

Ideally, we would like to address such questions by reconstructing the
evolutionary path followed by CVs empirically. In practice, this means
that we want observations to tell us how the secular mass-transfer
rate in CVs depends on orbital period. The word ``secular'' is key
here, since it encapsulates the main difficulty in this project. The
problem is that most conventional tracers of $\dot{M}$ -- in
particular those tied to the accretion luminosity -- are
necessarily measures of the {\em instantaneous} mass transfer rate in
the system. However, from an evolutionary perspective, what we need is
the secular accretion rate, i.e. $\dot{M}$ averaged over evolutionary
time-scales. The trouble is that there is no guarantee that
instantaneous and long-term $\dot{M}$ are the same. In fact, it has
been known for a long time that CVs with apparently very different
instantaneous accretion rates (e.g. dwarf novae and nova-likes) can
co-exist at the same orbital periods. One possible explanation is that
CVs may undergo irradiation-driven mass-transfer cycles on time-scales
of $10^5$~yrs (the thermal time-scale of the donor's
envelope; e.g. \cite{buning2004}).  

Recent years have seen the emergence of {\em two} new methods to overcome
this problem. The first is based on the 
properties of the accreting WDs in CVs, the second on the properties
of their mass-losing donors. The WD-based method builds on the
theoretical work of Dean Townsley and Lars Bildsten, who have shown 
that the (quiescent) effective temperature of an accreting WD in a CV
is a tracer of $\dot{M}$ \cite{townsley2002,townsley2003}. The 
donor-based method, on the other hand, exploits the fact that CV
secondaries are driven out of thermal equilibrium, and hence inflated, by
mass loss (see Figure~\ref{fig:mr}). This makes it possible to use the
degree of donor inflation as a tracer of secular
$\dot{M}$ \cite{knigge2010} (see also \cite{sirotkin2010}).

Both methods have their drawbacks, of course. WD-based $\dot{M}$
estimates are sensitive to the masses of the WD and its accreted
envelope (which are usually not well known), plus there remains a
residual $T_{eff}$ response to long-term $\dot{M}$ variations,
especially above the period gap. The main weaknesses of the 
donor-based method are its strong reliance on theoretical models of
low-mass stars, as well as its sensitivity to 
apparent donor inflation unrelated to mass loss (e.g. due to 
tidal/rotational deformation, or simply as a result of model
inadequacies).

The first results obtained by the two methods are shown 
in Figure~\ref{fig:reconstruct}. The left panel is from work by 
Dean Townsley and Boris
G\"{a}nsicke \cite{townsley2009} and shows how
$T_{eff}(P_{orb})$ predicted 
by different evolutionary models (including the standard one) compare
to a carefully compiled set of observed WD temperatures. The right
panel is from work by Isabelle Baraffe, Joe
Patterson and myself \cite{knigge2010} and shows a similar comparison
between (standard and non-standard) models and data in the donor
mass-radius plane.  

A full discussion of these results would go far beyond the scope of
this brief review, so I will focus on just one important aspect. Taken
at face value, both methods seem to suggest that GR alone is not
sufficient to drive the observed mass-loss rates below the period
gap. However, much work remains to be done in testing these methods,
exploring their limitations, verifying such findings and studying
their implications.\footnote{Indeed, Littlefair et
al. \cite{littlefair2008} have already suggested
that WD temperatures in a sub-sample of short-period CVs with
well-constrained WD masses {\em are}, in fact, consistent with purely
GR-driven AML. Sirotkin \& Kim \cite{sirotkin2010} have made the
same claim using a donor-based method (although their $\dot{M}$
estimates are based on highly simplified stellar models).}
What is clear, however, is that we finally have the tools to test 
the standard model quantitatively and, if necessary, to derive an
empirically-calibrated alternative model that can be used as a
benchmark in population synthesis and other studies. In fact, the
best-fit donor-based model in the right panel of
Figure~\ref{fig:reconstruct} is intended to provide exactly such an
alternative (see Ref. \cite{knigge2010} for details).

\begin{figure}[t]
  \includegraphics[height=.25\textheight]{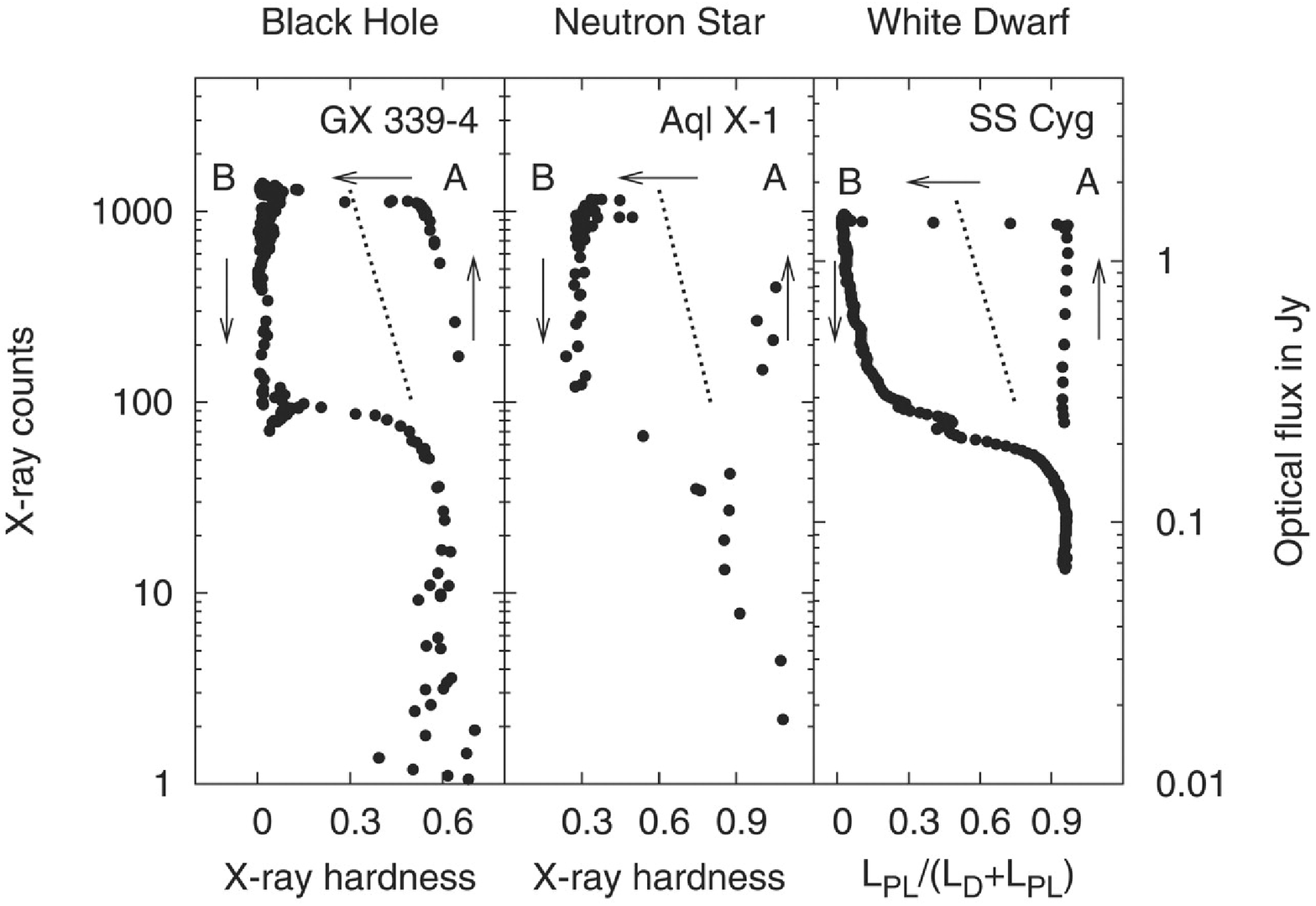}
  \includegraphics[height=.25\textheight]{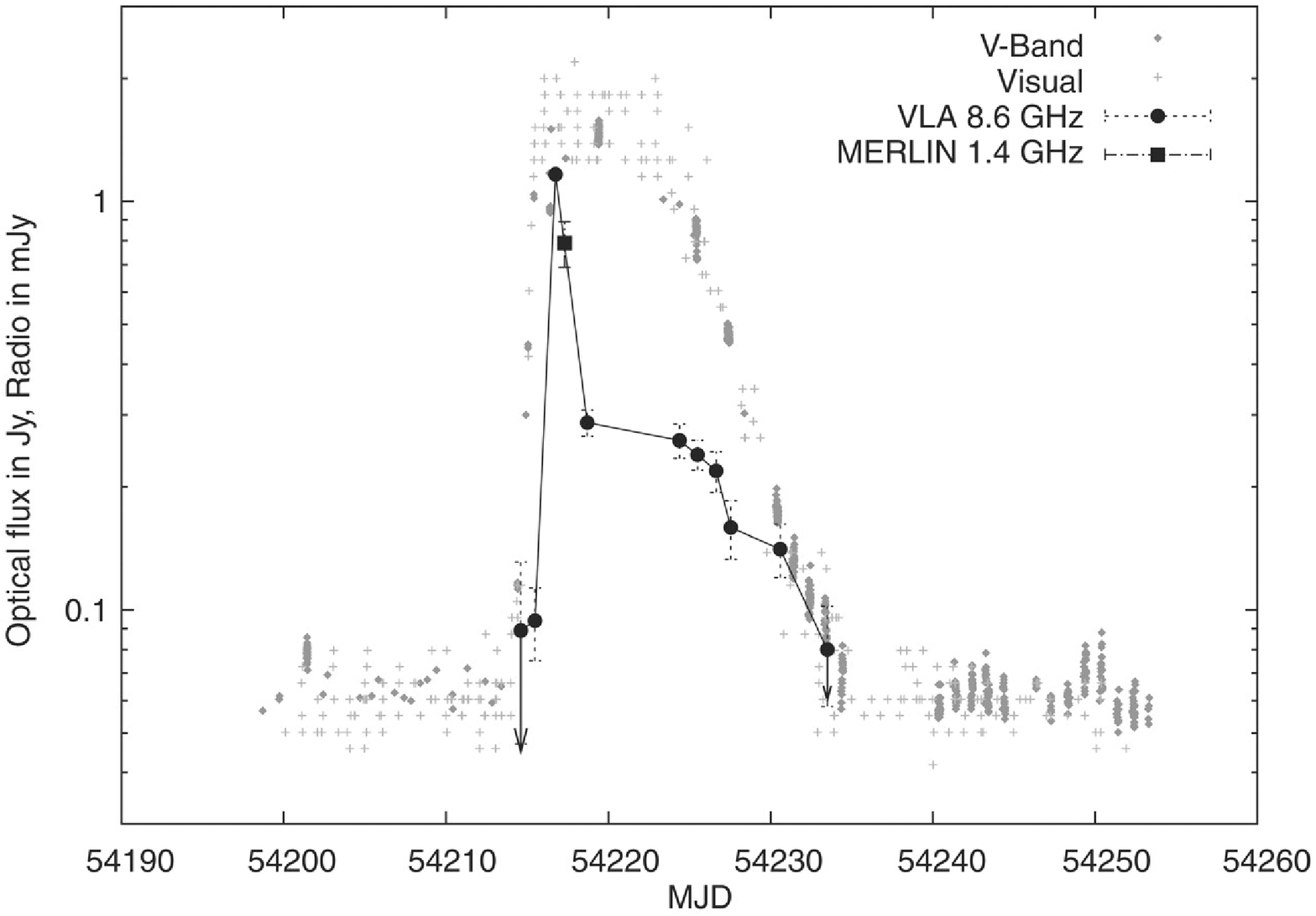}
  \caption{{\em Left Panel:}
  Radio and optical light-curve of the dwarf nova SS Cyg. Note the
  strong radio flare associated with the transition from quiescence to
  outburst. Figure reproduced from \cite{koerding2008}.
  {\em Right Panel:} Hardness-intensity diagram for a black hole, a
  neutron star, and SS Cyg. The arrows indicate the temporal
  evolution of an outburst. The dotted lines indicate the ``jet line''
  observed in black hole and neutron star XRBs. On its right side, one
  generally observes a compact jet; the crossing of this line usually
  coincides with a radio flare. For SS Cyg, a disc-fraction
  luminosity diagram is shown, i.e. optical flux is plotted against
  the power-law 
  fraction measuring the prominence of the "power-law component" in
  the hard x-ray emission in relation to the boundary layer/accretion
  disk luminosity. This power-law fraction has similar properties to
  the X-ray hardness used for XRBs. Figure reproduced by permission
  from Ref. \cite{koerding2008}.}
\label{fig:jets}
\end{figure}

\section{Breakthrough VI: The discovery of radio jets in CVs}

One of the interesting and counterintuitive aspects of accretion
physics on all scales -- from young stellar objects, CVs and 
low-mass X-ray binaries (LMXBs) all the way to active galactic nuclei
and quasars -- is that disk accretion is very often accompanied by some
form of bipolar outflow. In CVs, it has has long been known that
systems characterized by relatively high accretion rates produce
weakly collimated accretion disk winds. The evidence for these winds
comes primarily from the classic P-Cygni line profiles they display in
their ultraviolet spectra (e.g. \cite{cordova1982, long2002}).
However, many other types of accreting
objects (also) produce highly collimated {\em jets}, which had never 
been observed in CVs. Until recently, it was quite unclear whether
this absence of evidence for jets in CVs meant that they were hard to
find (e.g. \cite{knigge1998}), or that we were searching in the wrong
way, or whether it actually meant that CVs were missing a necessary
ingredient for jet formation, such as a powerful central energy source
(e.g. \cite{livio1999}).

This question, too, has finally been answered. The crucial insight
was provided by Elmar K\"{o}rding, who used the 
well-established jet phenomenology in BH and NS LMXBs
\cite{fender2004} to predict the optimal way to detect radio jets in
CVs, if they existed. The key point is that jet power in LMXBs
initially increases 
with $\dot{M}$, but is eventually quenched, with the quenching often
being preceded by bright radio flares. Thus, counterintuitively, the
best targets for a CV jet survey are not the steady high-$\dot{M}$
systems. but nearby dwarf novae on the rise from quiescence to
outburst.

The very first observational campaign designed to exploit this
predicted behaviour was successful. As shown in Figure~\ref{fig:jets}, with
the help 
of the AAVSO, we (i.e. K\"{o}rding et al., Ref. \cite{koerding2008}) 
caught the dwarf nova SS Cyg very early on the rise
to outburst. Exactly as predicted, this rise was  accompanied by a
sharp radio flare, whose properties are completely consistent with
those expected for a radio jet. In addition, we also 
compared the overall temporal evolution of outbursts
in CVs to those in BH and NS LMXBs and showed that they display
exactly the same type of hysteretic behaviour in what might be called
the ``colour-magnitude diagram'' of such outbursts
(Figure~\ref{fig:jets}). Thus not only are CVs capable of driving jets, but 
the entire unstable accretion process operating in CVs and LMXBs
appears to be similar in both classes of objects. One immediate
implication is that theoretical models of jet formation that rely on
ultra-strong gravitational or magnetic fields near NSs or BHs are
ruled out. 

\begin{figure}[t]
  \includegraphics[height=.35\textheight]{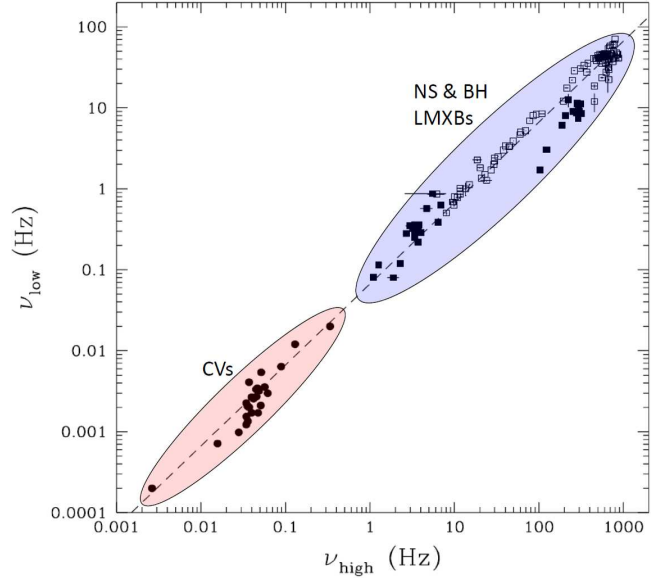}
  \caption{The two-QPO period diagram for X-ray binaries (filled
  squares: BH binaries; open squares: NS binaries) and 26 CVs (filled
  circles). Each CV is plotted only once. The X-ray binary data are
  from \cite{belloni2002}. The dashed line marks
  $\nu_{DNO}/\nu_{QPO} = 15$. Figure adapted and reproduced by
  permission from Ref. \cite{ww_summ}.}
\label{fig:qpo}
\end{figure}

\section{Breakthrough VII: Time Variability and Oscillations}

Another observational feature that appears to be common to accreting
systems of of all types and on all scales is short time-scale
variability in the form of stochastic flickering and/or (quasi-)periodic
oscillations. The origin of these oscillations is, in general, still
poorly understood, but it is clear that they are closely connected to
the accretion and outflow processes in the innermost disk
regions \cite{warner2004review,belloni2010}.

It has been known for some time that, in LMXBs, many, if not all, of
the observed periodic and quasi-periodic oscillations are
correlated (e.g. \cite{psaltis1999,belloni2002}). A particularly clean
correlation exists 
between the so-called ``lower kilo-Hertz oscillation'' (LKHO)and the
``upper horizontal branch oscillation'' (UHBO), with the former always
being characterized by a frequency that is $\simeq 15$-times that of the
latter.

In 2002, Brian Warner and Patrick Woudt pointed out that at least
one CV, the dwarf nova VW Hyi, seemed to produce an analogous 
pair of oscillations \cite{warner2002}. In
CVs, the two frequencies in question are called ``dwarf nova
oscillations'' (DNOs) and ``quasi-period oscillations'' (QPOs), and 
Chris Mauche soon provided another example of a system with
the same ratio between DNO and QPO frequencies \cite{mauche2002}. 
Since then, Warner, Woudt and Magaretha Pretorius have steadily
increased the number of CVs with measured DNO and QPO periods 
\cite{ww1,ww2,ww3,ww4,ww5,ww6}.

A recent compilation of their
estimates, along with the corresponding estimate for a sample of
LMXBs, is shown in Figure~\ref{fig:qpo}.  It is obvious that the
originally suspected trend continues to hold. Thus the data shown
includes  DNO/QPO pairs for 26 CVs, all of which lie along a
well-defined extension of the LKHO/UHBO relationship for LMXBs. 

How do we know that this is more than just numerology? A crucial point
is that the periods of LKHOs in LMXBs, as well as those of the DNOs in
CVs, are consistent with the respective dynamical time-scales at the inner
edges of the accretion disks in these systems. Thus there is a
physical reason to think that DNOs and LKHOs, and hence also UHBOs and
QPOs, are related by a common physical mechanism. If so, theoretical
models for these oscillations that rely on ultra-strong gravitational or
magnetic fields are again ruled out. More generally, the key point
is that the universality of accretion processes appears to extend not
just to outflows and jets, but also to variability. 

\section{Breakthrough VIII: Do All CVs Go Nova?}

\begin{figure}[t]
  \includegraphics[height=.32\textheight]{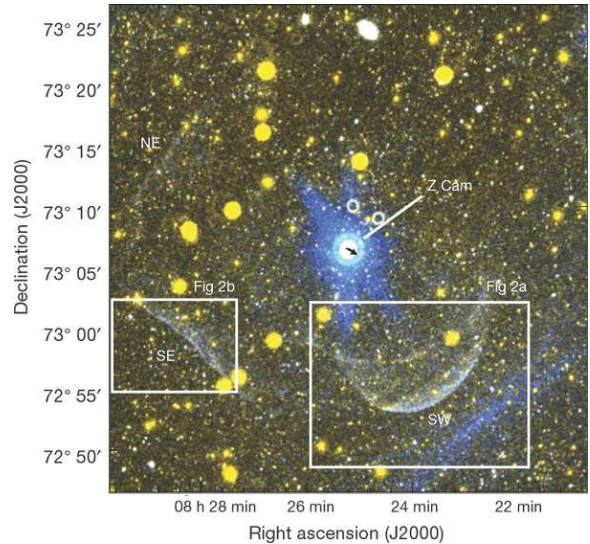}
  \caption{
  GALEX ultraviolet image of the field of the dwarf
  nova Z Cam (the bright star in the centre). Features associated with
  the nova shell around this object are highlighted by boxes. 
  Figure reproduced by permission from 
  Ref. \cite{shara2007}.}
\label{fig:nova}
\end{figure}

The final result I want to highlight concerns another piece of
long-standing CV lore. Every self-respecting CV researcher
``knows'' that nova eruptions are just a normal phase in the
life-cycle of all CVs. At first sight, this seems like an unassailable
proposition. After all, it is predicted theoretically 
(e.g. \cite{yaron2005}) and, observationally, it is well established
that, away from eruption, novae are basically just ordinary CVs
(e.g. \cite{warner_novae}).

Unfortunately, there is gaping hole in this logic: even if we allow
that all novae are CVs, this does not imply that all CVs eventually
become  novae. In principle, it would be perfectly possible that the nova
phenomenon is limited to a (possibly rare) sub-population of CVs.
Now it is, of course, impossible to prove that all CVs undergo
nova eruptions (not least because the theoretically expected nova
recurrence time-scales are typically $10^{4}-10^{5}$ years). However, it
would be immensely reassuring if there was even {\em one} CV that was
not actually {\em discovered} as a nova, but subsequently found to be
one. Even this is clearly difficult: we currently know $\sim 1000$ CVs, so
even if all of them were being monitored carefully, we could expect to
detect only one outburst every 10-100 yrs. These numbers
are extremely rough, since the predicted recurrence time scale depends
on the $M_{WD}$ and $\dot{M}$ (e.g. \cite{yaron2005}).

In 2007, Mike Shara and collaborators showed that there is indeed such
a system: the well-known dwarf nova Z Cam \cite{shara2007}. But this
was not a case of simply 
being lucky enough to catch the CV going nova. Instead, they
succeeded by discovering an ancient nova shell 
around this system (Figure~\ref{fig:nova}). These shells are composed
of material ejected in the eruption and can remain visible for $\sim
10^3$~yrs. This makes them an excellent tool for identifying CVs that
{\em used to be} novae. Thanks to this tool, we now know that (at
least some) ``ordinary'' CVs do, in fact, undergo nova eruptions.

There is an interesting postscript to this story. Shortly after the
publication of Ref. \cite{shara2007} in {\em Nature}, G\"{o}ran
Johansson \cite{johansson2007} pointed out in a letter that 
ancient Chinese documents analyzed by P.Y. Ho in 1962 \cite{ho1962}
report the appearance of a ``guest star'' near the location of Z Cam
in 77 BC. So, actually, the system {\em had} originally been
discovered as a nova... but was then lost again for over two
millenia.


\begin{theacknowledgments}
I would like to thank the conference organizers, particularly Vicky
Kalogera, for making this such an enjoyable meeting. Also, congratulations
again to Ron Webbink, whose 65$^{th}$ birthday provided the occasion
for this conference.
\end{theacknowledgments}



\bibliographystyle{aipproc}   


\bibliography{knigge}

\IfFileExists{\jobname.bbl}{}
 {\typeout{}
  \typeout{******************************************}
  \typeout{** Please run "bibtex \jobname" to optain}
  \typeout{** the bibliography and then re-run LaTeX}
  \typeout{** twice to fix the references!}
  \typeout{******************************************}
  \typeout{}
 }

\end{document}